\documentclass[article]{jss}
\usepackage{amsmath}
\usepackage{xcolor}
\usepackage{listings}
\usepackage{multicol}
\usepackage{xspace}

\newcommand\relem{\pkg{RElem}\xspace}
\newcommand\elemental{\pkg{Elemental}\xspace}
\newcommand\cpp{\proglang{C++}\xspace}
\newcommand\C{\proglang C\xspace}
\newcommand\R{\proglang R\xspace}

\definecolor{orange}{rgb}{1,0.5,0}
\definecolor{dgreen}{rgb}{0,0.5,0}

\lstdefinestyle{ccode}{language=C}
\lstdefinestyle{cppcode}{language=C++}
\lstdefinestyle{rcode}{language=R}
\lstdefinestyle{shell}{language=sh}

\author{Rodrigo Canales\\RWTH Aachen University \And 
    Elmar Peise\\RWTH Aachen University \AND
    Paolo Bientinesi\\RWTH Aachen University}
\title{Large Scale Parallel Computations in R through Elemental}

\Plainauthor{Rodrigo Canales, Elmar Peise, Paolo Bientinesi} 
\Plaintitle{RElem: Large Scale Parallel Computations in R through Elemental} 
\Shorttitle{RElem} 

\Abstract{
    Even though in recent years the scale of statistical analysis problems has
    increased tremendously, many statistical software tools are still limited to
    single-node computations.  However, statistical analyses are largely based
    on dense linear algebra operations, which have been deeply studied,
    optimized and parallelized in the high-performance-computing community.  To
    make high-performance distributed computations available for statistical
    analysis, and thus enable large scale statistical computations, we introduce
    \relem, an open source package that integrates the distributed dense linear
    algebra library \elemental into~\R.  While on the one hand, \relem provides
    direct wrappers of \elemental's routines, on the other hand, it overloads
    various operators and functions to provide an entirely native~\R experience
    for distributed computations.  We showcase how simple it is to port existing
    \R~programs to \relem and demonstrate that \relem indeed allows to scale
    beyond the single-node limitation of~\R with the full performance of
    \elemental without any overhead.
}
\Plainkeywords{linear algebra, elemental, distributed, parallel, R} 

\Address{
      Rodrigo Canales\\
      Aachen Institute for Advanced Study in Computational Engineering Science\\
      RWTH Aachen University\\
      Schinkelstr. 2\\
      52062 Aachen, Germany\\
      E-mail: \email{canales@aices.rwth-aachen.de}\\
      URL: \url{http://hpac.rwth-aachen.de}
}

\begin{document}

\section{Introduction}

In many scientific fields ranging from Chemistry and Biology to Meteorology,
data-sets of increasing size are analyzed with statistical methods on a daily
basis.  Such analyses are often performed with the help of~\R, an open-source
software and programming language focused on and designed for statistical
computations.  \R~is supported by both a large community and CRAN, a
well-developed contribution system with over 9000 open-source packages.
\R's~user-base has steadily grown since its first release in~1993 and in many
disciplines it has become the de-facto standard for statistical computations.
However, statistical analyses are constantly evolving, most notably, in the size
of their databases.  In fields like computational biology, it is quite common
that analyses are performed on workstations, and take days to complete.  Other
problems are simply too large to fit in even modern main memories.

In the field of scientific computing, which is often only loosely connected with
the statistical computing community, large scale problems are routinely solved
through distributed parallel computing:  In clusters, the combined resources and
processing power of multiple interconnected computers make processing large
data-sets feasible and fast.  For dense linear algebra operations, which are at
the core of statistical computing, there exist highly optimized libraries for
both shared-memory and distributed systems.

While \R already makes use of such optimized libraries, namely 
\pkg{BLAS}\footnote{Basic Linear Algebra Subprograms} and 
\pkg{LAPACK}\footnote{Linear Algebra PACKage} , for single-node
computations, it cannot profit from modern distributed memory libraries for
larger data-sets.  In this paper, we overcome this limitation by integrating
\R~with the state-of-the-art distributed linear algebra library \elemental:  We
propose the \R-package \relem.  \relem provides the functionality and
performance of \elemental in~\R and thereby enables large scale computations in
native~\R.  \relem not only provides wrappers to \elemental, but replicates
\R's~operators and methods, allowing the use of native \R~scripts in distributed
computing nodes.

The remainder of this paper is structured as follows:  In Sec.~\ref{sec:basics}
we summarize currently available distributed computing packages for~\R and
provide an overview of the distributed \elemental library.~\ref{sec:basics}.  In
Sec.~\ref{sec:relem} we then present how \relem connects \R~and \elemental in
and describe its usage in Sec.~\ref{sec:usage}.  Finally, in Sec.~\ref{perf} we
demonstrate the performance and large-scale capabilities of \relem and, in
Sec.~\ref{conclusions} draw conclusions and potential future directions.

\section{Distributed parallelism in R}
\label{sec:basics}

With the advent of multi-core processors, \R~began to embrace shared memory
parallelism.  Several built-in routines use multi-threading capabilities or rely
on parallel high-performance kernels, such as the \pkg{BLAS}.  On the
contrary, distributed parallelism is not natively available in~\R.

There are several packages that address this issue and enable multi-node
execution in~\R.  Most of these packages are built on top of the Message Passing
Interface (\pkg{MPI}), which enables the communication between different
processes.  \pkg{MPI} enables the Single Program Multiple Data (SPMD) parallel
programming paradigm, where all computing nodes execute the same program using a
portion of the global data.  Processes synchronize with each other using
messages, which can be point-to-point or collective among processes.

A low-level interface to this parallelism model in~\R is provided by \pkg{RMPI},
a package that exposes \pkg{MPI} routines in~\R but leaves the parallelization
task to the user.

Other packages such as \pkg{Snow} and \pkg{Snowfall} apply one function to or
perform reductions on a whole distributed data set in parallel.  These packages
achieve good performance and their use is straightforward for simple
parallelization tasks such as numerical integration~\citep{parallelR}.

The package collection \textit{Programming Big Data in~\R} (\pkg{pbdR)}
\citep{pbdR} provides mechanisms and routines to handle big amounts of data in
distributed environments.  In particular, the module \pkg{pdbDMAT} offers
distributed dense linear algebra functionality based on the \pkg{ScaLAPACK}
library (a distributed memory parallelization of \pkg{LAPACK})
\citep{scalapack}.  It also implements native \R~interfaces to and operators on
distributed matrices, making it accessible without extensive knowledge of
distributed computing in general or \pkg{ScaLAPACK} in particular.

However, in the field of parallel dense linear algebra, \pkg{ScaLAPACK} (1992)
is nowadays seen as a legacy library that is still widely used, but no longer
represents the state of the art.  For instance, it does not contain more recent
developments on distributed linear algebra for key routines, such as eigenvalue
decomposition \citep{gutheil2014performance}.  To provide \R~users access to the
state of the art in dense linear algebra, we therefore see the necessity to
provide a more newer alternative.  For this purpose, we use \elemental
\citep{Elemental}, a modern distributed linear algebra library that incorporates
the latest algorithms and modern programming standards.

\elemental is a recent (2013) and actively developed library.  Since its open
source license (BSD) allows it to be included in proprietary software, it is an
ideal basis for distributed computations both in academia and industry.

In contrast to legacy libraries such as \pkg{BLAS}, \pkg{LAPACK}, and
\pkg{ScaLAPACK}, \elemental embraces modern programming paradigms to provide a
simpler object-oriented programming interface.  At the center of this interface
are distributed matrices, which encapsulate not only their data but also their
size and distributed storage format.  All of \elemental's computational and
input/output routines implicitly operate on these formats, performing optimized
matrix communications, replications, and redistributions as necessary, while
keeping the memory overhead minimal.  Hence, while the distributed storage
formats are fully transparent and accessible by the user, knowledge of such is
not required to use the library.

To extend the availability of \elemental beyond its native \cpp~environment,
recent releases include \C~and \proglang{Python} interfaces that cover the
majority of the library's functionality.  To also make the powerful
functionality and features of \elemental available to the statistical
programming community, we embrace the task of integrating the library with~\R.

\section{RElem}
\label{sec:relem}

This section introduces \relem, our novel interface package that provides
\elemental features and functionality in native~\R.  We begin with an overview
of \relem's design and structure and in the following subsections focus on its
implementation.

The \R~programming language is in part so widely used in statistical computing
and data analysis not so much because of its inherent feature set but because of
its large and well maintained package system and simple extensibility.  It can
natively interface with \proglang{Fortran} and \C~routines, and separate
packages provide interaction with other languages such as \proglang{Java}
and~\cpp~\citep{Rcpp}.

Our first attempt to connect \R with \elemental was to use the interface code
generator SWIG \citep{BeazleySwig} to directly interface with \elemental's core
\cpp~interface.  However, we soon ran into complications due to the extensive
use of templating and \proglang{C++11} features within \elemental.  As an
alternative, we turned to \elemental's \C~interface, which also serves as the
basis for the library's \proglang{Python} interface.

With \C as a starting point, there are two interfacing mechanisms provided
by~\R:  On the one hand, the routine \code{.C} operates as a simple \C~wrapper
that reinterprets \R~data as basic \C~data types; on the other hand, the routine
\code{.Call} provides a low-level interface to compiled \C~object files.  As
arguments, \code{.Call} passes and operates on internal \R~objects, the
so-called SEXP \C~structures.

Since \elemental operates on matrices and other complex objects that are neither
basic \C~data-types nor known to~\R, the \code{.C} interface is not applicable
and we have to employ \R's~\code{.Call} interface.  This means that a separate
layer between \R's~internal SEXP objects and \elemental's \C~interface is
required. However, this also allows to directly benefit from much of
\R's~features, such as object methods, garbage collection, and type checking.  

\begin{figure}[h]
    \centering
    \includegraphics[width=0.7\linewidth]{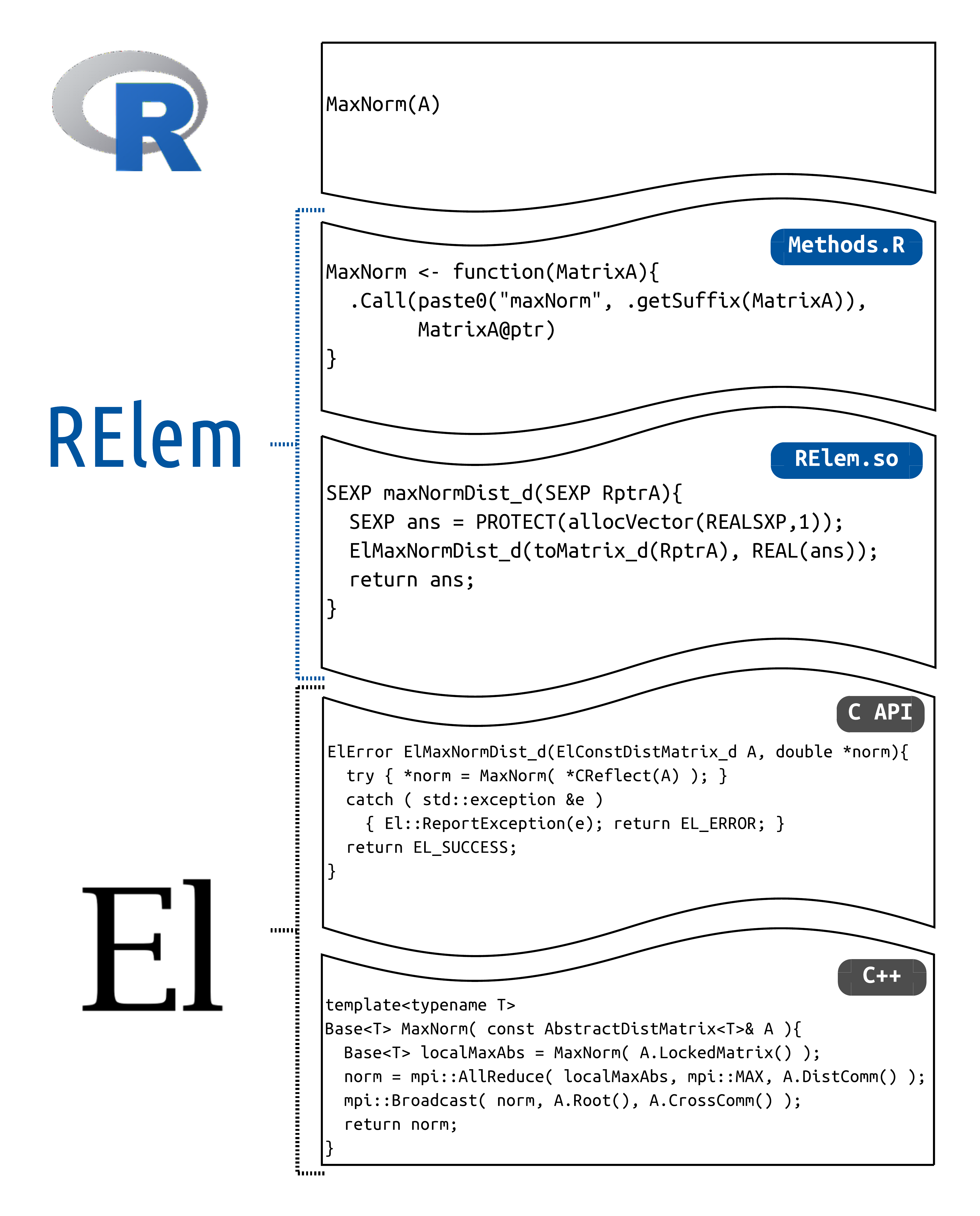}
    \caption{\relem's integration scheme.}
    \label{scheme}
\end{figure}

In the following, we present the layered structure of \relem from bottom to top
as depicted in Fig.~\ref{scheme}:  We begin with a short description of
\elemental's \C~interface and describe its mapping to \R's~internal SEXP format.
We then provide an overview of the \R~wrappers using the \code{.Call} interface:
The definition of \elemental matrix types and routines in~\R, as well as the
representation of \elemental's object-oriented methods of matrix classes.  We
conclude with the extension \elemental's functionality by overloading
\R~operators and functions to provide an interface very similar to native~\R.

\subsection{Elemental's C interface}
\label{ElCAPI}

To provide a simple interface to languages other than~\cpp, \elemental offers a
\C~compatibility API  and headers.  In this API, \elemental's templated class
system is translated into fully typed \C~structures and functions:
\begin{itemize}
    \item The templated classes are instantiated with \code{El} as a prefix
        (corresponding to the \cpp namespace \code{El}), and suffixes indicating
        the data-types, e.g., \code{ElMatrix_d} for a local double-precision
        matrix.  The instantiations are implemented as anonymous structures that
        are essentially pointers to the \cpp~objects.
    \item Templated routines are similarly instantiated, e.g., the \cpp~function
        \code{MaxNorm(A)} in~\C becomes \code{ElMaxNormDist_c(A, \&val)} for
        distributed single-precision complex matrix.  \cpp~return values are
        implemented as the last argument by reference, and each routine returns
        an error code representing errors thrown in the corresponding
        \cpp~routine.
    \item Instance methods are implemented as routines that take a reference to
        the instance as a first argument; e.g., \code{A.Width()} becomes
        \code{ElMatrixWidth_i(A)} for local integer matrices.
\end{itemize}

\subsection{RElem's C-layer}
\label{RElemCLayer}

Since \relem relies on \R's~\code{.Call} routine to interact with \elemental,
the construction of a \C-layer that connects \R's~SEXP structures and
\elemental's \C~interface is necessary.  This layer contains a wrapper for each
of \elemental's \C-routines.  In these wrappers, we already simplify some
abstractions from the \C~interface by implementing return values as such,
instead of final arguments by reference.  

\begin{lstlisting}[
    style=ccode, label={RElemC},
    caption={\code{maxNormDist\_d} in \relem's \C-layer.}
]
SEXP maxNormDist_d(SEXP RptrA){
    SEXP ans = PROTECT(allocVector(REALSXP, 1));
    ElMaxNormDist_d(toDistMatrix_d(RptrA), REAL(ans));
    UNPROTECT(1);
    return ans;
}
\end{lstlisting}

As an example, listing~\ref{RElemC} shows the definition for
\code{maxNormDist_d} that serves as a wrapper for \elemental's
\code{ElMaxNormDist_d}.  The routine begins by allocating a double-precision
return value using \R's~internal allocator and types\footnote{The \code{PROTECT}
and \code{UNPROTECT} are necessary to indicate to \R's~garbage collection that
the allocated object is in use.}.  It then invokes \elemental's \C~interface,
passing arguments extracted from the SEXP objects:  \code{toDistMatrix_d} is an
\relem routine that casts the external pointer stored in a SEXP object to a
\code{ElDistMatrix_d}, and \code{REAL} is an \R~macro that provides a pointer to
the double precision slot in a SEXP.  Finally, the SEXP representing the result
is returned.

\subsection{R Matrix and Function Wrappers} 
\label{RElemWrappers}

The \C-layer described above allows to load and access \elemental's \C~interface
in~\R through the \code{.Call} syntax.  In order to offer not a mere copy of
\elemental's \C~interface but mirror its original object-oriented
\cpp~functionality in~\R, \relem contains wrappers for all \elemental data types
(e.g., matrices) and routines (e.g., \code{MaxNorm}).

Just as \elemental, \relem represents both local and distributed matrices (as
well as few other objects) as classes\footnote{We use \R's~recommended S4~object
system.}.  Each of these classes contains 1) an external pointer to an
\elemental object, and 2) a datatype indicator needed to select the
corresponding typed \C-layer routines.

\begin{lstlisting}[
    style=rcode, label={DMat},
    caption={R class \code{ElDistMatrix}.}
]
ElDistMatrix <- setClass(
    "ElDistMatrix",
    representation(
        ptr      = "externalptr",
        datatype = "character"
    )
)
\end{lstlisting}

As an example, listing~\ref{DMat} presents the definition of the class
\code{ElDistMatrix}.  In addition to the class name, the class declaration
routine \code{SetClass} is passed the class ``representation'' containing its
attributes:  The external pointer \code{ptr} references the \elemental matrix as
a \C~pointer and the character \code{datatype} indicates the type of data in
this matrix, i.e., double-precision real (\code d) or complex (\code z), or
integer-precision (\code i).  Note that, in contrast to \elemental's
\C~interface, which has separate structures and routines for each data-type,
thanks to this mechanism \relem has only one distributed matrix class that
encapsulates the data-type as an attribute.

\begin{lstlisting}[
    style=rcode, label={MaxNorm},
    caption={\R wrapper for \code{MaxNorm}.}
]
MaxNorm <- function(MatrixA) {
    .Call(
        paste0("maxNorm", .getSuffix(MatrixA)),
        MatrixA@ptr
    )
}
\end{lstlisting}

As shown in the example of \code{MaxNorm} in listing~\ref{MaxNorm}, To translate
\elemental functions on matrices to~\R, we need to use the \code{.Call}
interface.  The name of the function called in \relem's \C-layer is constructed
from two parts:  the name \code{MaxNorm} and a type-dependent suffix that is
constructed from the information encapsulated in the matrix object by
\code{.getSuffix}.  E.g., for a complex distributed matrix, the result would be
\code{MaxNormDist_z}, while for a real local matrix, it would be
\code{MaxNorm_d}.  The selected \C-function is then invoked with the external
pointer to the \elemental matrix (\code{MatrixA@ptr}) as an argument.  The
return value of the \C-function and thus \code{.Call} is implicitly returned by
the \R~wrapper.

\begin{lstlisting}[
 style=rcode, label={initDMat},
 caption={\code{DistMatrix} constructor.}
]
setMethod(
    "initialize",
    signature(.Object="ElDistMatrix"),
    function(.Object, tag="d", colDist="MC", rowDist="MR",
             grid=new("ElGrid")) {
        .Object@datatype <- tag
        .Object@ptr <- .Call(
            paste0("newDistMatrixSpecific_", tag),
            colDist, rowDist, grid@ptr
        )
        reg.finalizer(.Object@ptr,
                      .ElDestroy(tag, "destroyDistMatrix"))
        return(.Object)
    }
)
\end{lstlisting}

Class constructors are implemented by overloading \R's~\code{initialize}
function as shown for \code{DistMatrix} in listing~\ref{initDMat}.  Of the
constructor's four arguments, the first (\code{tag}) indicates the data type and
determines the correct \C-layer constructor, while the remaining three relate to
\elemental's data distribution and are passed directly to this constructor and
further to \elemental's \C~interface.  However, all arguments are optional:  By
default, matrices use double-precision (\code d) as a data-type and the
remaining arguments map to the default \cpp~argument and template values.  The
\elemental object is initialized through the \C-layer constructor and the
corresponding \C-layer destructor is associated to the instance to integrate
with \R's~garbage collection.  Finally the newly created instance is returned.

\subsection{Matrix Methods}
\label{RElemMethods}

Methods on matrix instances, such as \code{A.Height()}, which yields the matrix
height, require a slightly different treatment from functions that operate on
such matrices as arguments.  What makes the issue surprisingly complicated (from
a programming language point of view) is that \R~does not have a direct
equivalent to instance methods and, in particular, the dot~\code. is not a
reserved symbol in~\R, i.e., it can appear as part of a variable name.  To
overcome this shortcoming, we use the same approach as many other \R~libraries
and resort to the dollar character~\code\$ as a substitute for the dot~\code. in
\cpp.  Technically, \code\$~is a binary operator in~\R that takes the left
and right hand side as arguments (such as \code+), i.e., \code{A\$arg} is
equivalent to
\code{\$(A, arg)}.

\begin{lstlisting}[
    style=rcode, label={DollarMethod}, 
    caption={Method accessor \code\$ for \code{ElDistMatrix}.}
]
setMethod(
    "$",
    signature(x="ElDistMatrix"),
    function(x, name, ...) {
        id <- pmatch(name, names(DistMatrixFunctions))
        if (is.na(id)) {
            stop(paste("function", name, "not found\n"))
        }
        routine <- DistMatrixFunctions[[id]]
        function(...) {
            routine(x, ...)
        }
    }
)
\end{lstlisting}

\begin{lstlisting}[
    style=rcode, label={ListFunc},
    caption={Mapping of instance methods to functions for \code{ElDistMatrix} (excerpt).}
]
DistMatrixFunctions <- list(
    "Get"    = MatrixGet,
    "Set"    = MatrixSet,
    "Height" = MatrixHeight,
    "Width"  = MatrixWidth,
    "LDim"   = MatrixLDim,
    "Empty"  = MatrixEmpty,
    ...
)
\end{lstlisting}

Listing~\ref{DollarMethod} shows how \code{setMethod}\footnote{\code{setMethod}
takes three arguments: a method name, a signature specifying the arguments and
their types, and a function.} overloads the operator~\code\$ for the class
\code{ElDistMatrix}:  The associated function receives as arguments both the
matrix instance and the name of the method.  It first looks up the method name
in \code{DistMatrixFunctions} (listing~\ref{ListFunc}) returning an \R~function
(upon failure an error is thrown);  the matched function is implemented as part
of the \R-layer, as described in the previous section.  The operator function
for~\code\$ returns an anonymous function that wraps the matched function,
passing the matrix instance as a first argument and any additional arguments
(masked as~\code{...}).

\subsection{Operator and Function Overloading}
\label{RElemOverload}

While the classes, functions, and methods described above are already sufficient
to cover \elemental's full functionality, \relem further overloads several
\R~functions and operators to simplify many operations and provide a native
experience to \R~users.

\begin{lstlisting}[
    style=rcode, label={plusMethod},
    caption={Operator \code+ for \code{ElDistMatrix}.}
]
setMethod(
    "+",
    signature(e1="ElDistMatrix", e2="ElDistMatrix"),
    function(e1, e2) {
        if (e1@datatype != e2@datatype)
            stop("Matrices must have the same datatype")
        if (e1$Height() != e2$Height() && e1$Width() != e2$Width())
            stop("Matrices must have the same size")
        matA <- DistMatrix(e1@datatype)
        Copy(e1, matA)
        Axpy(1.0, e2, matA)
        matA
    }
)
\end{lstlisting}

The overloading arithmetic operators, such as \code+, \code{\%*\%}
and~\code*\footnote{matrix-matrix and element-wise multiplication.} through
\code{setMethod} is very similar to the already overloaded operator~\code{\$},
yet requires a few additional argument checks and a memory allocation for the
return value.  The overloading of~\code+ for \code{ElDistMatrix} is shown in
listing~\ref{plusMethod}:  The operator receives two distributed matrices;
first, it is ensured that both the matrix data-types and their sizes match; then
a new matrix is created and initialized with a copy of the first operand;
finally, the actual computation is performed by \code{Axpy}, an \elemental
function named after the \pkg{BLAS} counterpart that scales (in this case by
1.0) and adds one matrix to another; the new matrix is returned.

\begin{lstlisting}[
    style=rcode, label={indexDMat}, 
    caption={Indexing overloading in \R for \code{ElDistMatrix}.}
]
setMethod(
    "[",
    signature(x="ElDistMatrix"),
    function(x, i, j, ...) {
        if (length(i) == 1 && length(j) == 1)
            return(MatrixGet(x, i - 1, j - 1))
        g <- Grid()
        DistMatrixGrid(x, g)
        V <- DistMatrix(grid=g, tag=x@datatype);
        ViewNormal(V, x, i[1] - 1, tail(i, 1), j[1] - 1, tail(j, 1))
        V
    }
)
\end{lstlisting}

In~\R, arrays and tables con be indexed not only to access single elements but
to address portions of such objects; e.g., indexing allows to extract certain
rows or columns from matrices.  Since this is also a very common operation in
dense linear algebra, \elemental provides so-called ``matrix views''.  We use
this feature to overload \R's~indexing operator~\code[ as show for
\code{ElDistMatrix} in listing~\ref{indexDMat}:  If the provided indices \code i
and \code j identify a single matrix element, this is returned as a scalar;
otherwise, a submatrix is extracted.

\begin{lstlisting}[
    style=rcode, label={printDMat}, 
    caption={Overloading of \R's \code{print} for \code{ElDistMatrix}.}
]
setMethod(
    "print",
    signature(x="ElDistMatrix"),
    function(x) {
        Print(x)
    }
)
\end{lstlisting}

Native \R~methods are overloaded very similarly, as shown for \code{print} and
\code{ElDistMatrix} instances in listing~\ref{printDMat}: The overload simply
maps \R's~\code{print} to \relem's \code{Print} wrapper function.

\begin{lstlisting}[
    style=rcode, 
    caption={Overloading of \R's SVD-based PCA routine \code{prcomp} for
    \code{ElDistMatrix}.}, 
    label={prcomp}
]
setMethod(
    "prcomp",
    signature(x="ElDistMatrix"),
    function(x, retx=TRUE, center=TRUE, scale.=FALSE, 
             tol=NULL, ...) {
        x_centered <- scale(x, center, scale.)
        x_center   <- attr(x_centered, "scaled:center")
        s          <- svd(x_centered, nu=0)
        s$d        <- (1 / sqrt(x$Height() - 1)) * s$d
        list(sdev=s$d, rotation=s$v, center=x_center)
    }
)
\end{lstlisting}

Other core \R~operations, such as the principal component analysis (PCA), do not
have a direct counterpart in \elemental, but can be assembled from \elemental
routines.  For instance, the PCA routine \code{prcomp}, which is based on the
singular value decomposition (SVD) of the $n \times m$ dataset $X$ with $n$
observations of $m$ variables, is overloaded in \relem as outlined in
listing~\ref{prcomp}:  First, \code{scale}, which is also overloaded in \relem,
centers (and optionally scales) the input $X$ to ensure that the average of each
observed variable is~0: $\tilde X = X - \bar X$; the center $\bar X$ is
extracted since it is one of \code{prcomp}'s list of return values.  Now
\code{svd} performs the core singular value decomposition $U \Sigma V^T = \tilde
X$ through \elemental's \code{SVD}.  While the principle components are directly
given by the right singular vectors $V$ stored in \code{s\$d}, the variance
$\sigma$ for each vector is computed from the singular values (diagonal elements
of $\Sigma$), and then stored in \code{s\$v} as $\sigma = \frac\Sigma{\sqrt{n -
1}}$.  The returned data structure consists of the principle components, their
variance, and the matrix center.

\section{Usage}
\label{sec:usage}

This section offers a brief introduction to \relem from the user's perspective:
We start by describing \relem's installation process, then introduce its
parallel execution across multiple processes, and conclude by comparing
distributed \relem~scripts with native~\R, and with \elemental.

\subsection{Installation}

\relem is provided as a standard \R~package and installed through the command

\begin{lstlisting}[style=shell]
R CMD INSTALL RElem.zip
\end{lstlisting}

Since \elemental is not distributed and compiled along with \relem, the
installation script searches the file system for a valid \elemental installation
and compiles \relem accordingly using \elemental headers.  \relem has been
successfully installed and tested on both Linux and OS~X.\footnote{Windows is
currently not supported by \elemental but is also uncommon on distributed memory
cluster systems.}

\subsection{Parallel Launch}

Since \elemental is based on \pkg{MPI} and is typically launched in parallel
with \code{mpirun}, so is \relem:  \code{mpirun} is used to run one instance
of~\R per process, each of which receives the same input file.  \elemental is
loaded locally on each process and performs all interaction with \pkg{MPI} ---
the \R~instances are at no point confronted with communication and are actually
unaware that they are running in parallel.  A user might for instance run an
\relem script \code{myScript.R} on 64~processes using

\begin{lstlisting}[style=shell]
mpirun -n 64 R -f myScript.R
\end{lstlisting}

\subsection{Syntax and Interface}

The component to successfully run \relem is the \R~script that uses it.  To give
an impression not only of how \relem is used but also of how it compares to
pure~\R and \elemental, let us start with a simple example in pure~\R that loads
a Hermitian matrix from a file \code{data.txt}, and computes its
eigendecomposition, storing the result (eigenvalues and eigenvectors) in the
structure \code{ans}:

\begin{lstlisting}[
	style=rcode,
	caption={Pure \R eigen decomposition.},
	label={eigenR}]
A <- read.table("data.txt")
ans <- eigen(A)
\end{lstlisting}

To convert the above code to use \relem's distributed matrices, only few changes
are necessary:  

\begin{lstlisting}[
	style=rcode,
	caption={RElem eigen decomposition.},
	label={eigenRElem}]
library(Relem)
A <- read.table.dist("data.txt")
ans <- eigen(A)
\end{lstlisting}

Besides loading \relem, the script only needs the addition of \code{.dist} to
the \code{read.table} routine, which then loads the passed file as
\code{ElDistMatrix}.

If we compare the above \relem script with a \cpp~code performing the same
operation, we notice that the \relem version is considerably shorter:

\begin{lstlisting}[
    style=cppcode, 
	caption={Elemental C++ eigen decomposition.},
	label={eigenRElem}
]
#include "El.hpp"
using namespace El;
int main(int argc, char *argv[]) {
    Initialize(argc, argv);
    DistMatrix<double> A, w, X;
    Read(A, "data.txt");
    HermitianEig(LOWER, A, w, X);
    return 0;
}
\end{lstlisting}

\section{Performance and Scalability}
\label{perf}

In this chapter we present the performance and scalability of \relem; we compare
it with both pure~\R and \elemental and show computations only made possible
in~\R through \relem.  We demonstrate that \relem translates both \elemental's
performance and scalability directly to~\R with minimal overhead.

Our test environment is a computing cluster of nodes equipped with two 12-core
Intel Xeon E5-2680 processors (Haswell microarchitecture) and 64GB of main
memory.  We use \R~version~3.1.0, \elemental version~0.85, and link to Intel's
Math Kernel Library (\pkg{MKL}) version 11.3.

\subsection{Performance: R vs. RElem}

We begin with a performance comparison on a single node (24~cores) between
\R~linked to multi-threaded \pkg{MKL}, and \relem using \elemental's
\pkg{MPI}-based parallelization; as test-cases we use two basic linear algebra
operations:  the multiplication of two $6{,}000 \times 6{,}000$ matrices
(\code{gemm}, the central and most optimized operation in dense linear algebra)
and the solution of a linear system with $12{,}000$ unknowns and $1{,}000$
right-hand sides.

\begin{figure}[h]
    \centering
    \includegraphics[width=0.7\linewidth]{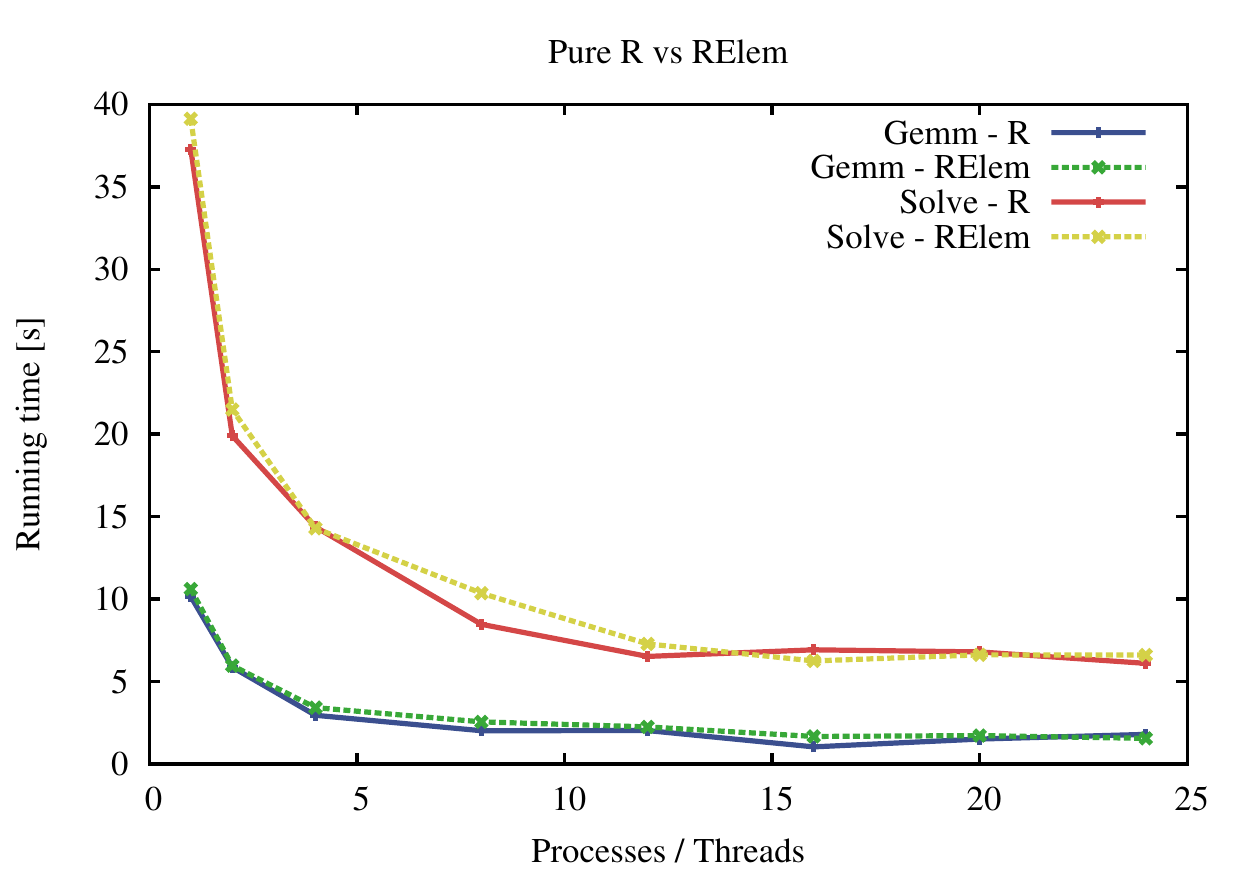}
    \caption{Execution time in \R and \relem.}
    \label{LapackPerformance}
\end{figure}

Figure~\ref{LapackPerformance} presents the execution time of these two
operations for both \R~and \relem, with increasing number of cores (threads
in~\R, processes in \relem):  Up to small differences, the libraries are on-par
with one another, as neither is clearly faster than the other.

Note that an alternative would be to run one instance of \relem per node and
rely on multi-threaded \pkg{mkl} for shared memory parallel performance.

\subsection{Scalability: R vs. RElem}

The main goal of \relem is to extend \R's~processing capabilities beyond the
memory and compute limitations of a single note.  Hence, in the following, we
consider a Principle Component Analysis (\code{prcomp}) with $5{,}000$ variables
and between $125{,}000$ and $875{,}000$ observations, i.e., an input matrix that
is between 4.66GB and 41.90GB.

\begin{figure}[h]
    \centering
    \includegraphics[width=0.7\linewidth]{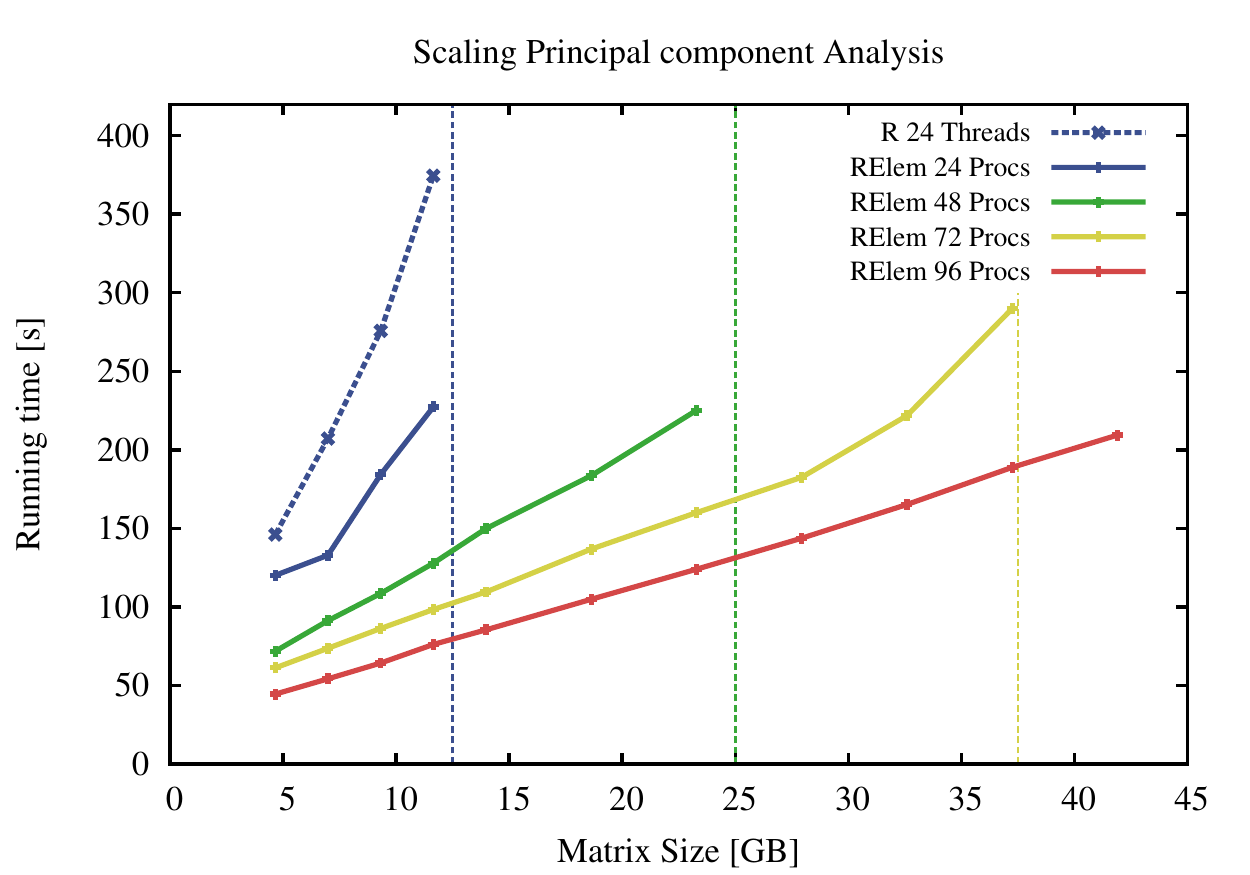}
    \caption{PCA Execution time in \R and \relem. Vertical lines: memory
    limitations per node.}
    \label{PCAScaling}
\end{figure}

Figure~\ref{PCAScaling} presents the execution time of the PCA of both \R~and
\relem for increasing problem sizes.  First off, already on a single node,
\relem is between 18\% and 43\% faster than native~\R since \elemental offers a
faster singular value decomposition algorithm than \pkg{MKL}.  Furthermore,
since during the PCA computation multiple copies of the input matrix are
created, both \R and \relem can only solve problems of up to 12GB in size
(roughly $320{,}000$ observations) on a single node.  While this limitation
cannot possibly be overcome in native~\R, it is easily addressed in \relem by
adding more processes.  In addition to speeding up single-node analyses, each
added compute node allows to linearly increase the size of the processed problem
by 12GB.

\subsection{Overhead over Elemental}

The previous experiments have shown the performance and scalability advantages
of \relem compared to~\R; in the following, we consider \relem as an interface
to \elemental and study its overhead in execution time and memory consumption.

We begin by analyzing the execution time overhead of \relem over \elemental: For
this purpose, we consider the multiplication of two matrices of size $12{,}000
\times 12{,}000$ and the solution of a linear system with $20000$ unknowns and
$1000$ right-hand sides (both problems take up around 3GB).

\begin{figure}[h!]
    \centering
    \includegraphics[width=0.65\linewidth]{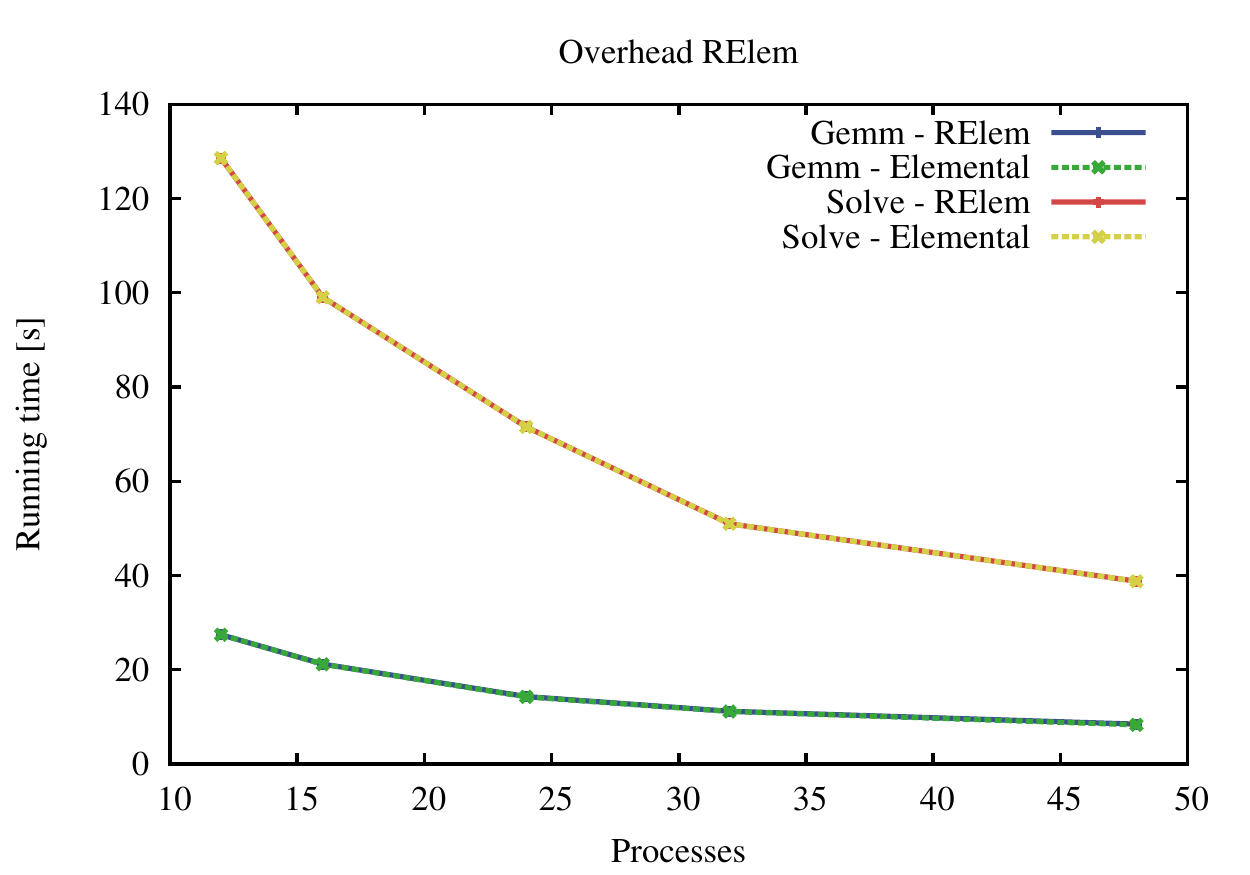}
    \caption{Performance comparison of \relem with \elemental}.
    \label{OverheadLAPACK}
\end{figure}

Figure~\ref{OverheadLAPACK} shows the execution time of these two operations for
both \relem and \elemental:  The two different setups are virtually identical,
showing that \relem causes no relevant overhead over \elemental for large
computations.  This behavior was expected since \relem only introduces very
light-weight wrapping layers around \elemental's core functions.  Measurements
of simple query routines such as \code{A.Width()} (In \relem: \code{A\$Width()})
confirm this observation and suggest a marginal overhead of, on average, 0.6ms
per function call.

Next, we study the memory overhead of \relem over \elemental; for this purpose,
we consider a principal component analysis with $5000$ variables and between
$125{,}000$ and $725{,}000$ observations (between 3.66GB and 27.00GB) on three
compute nodes (72 cores).

\begin{figure}[h!]
    \centering
    \includegraphics[width=0.65\linewidth]{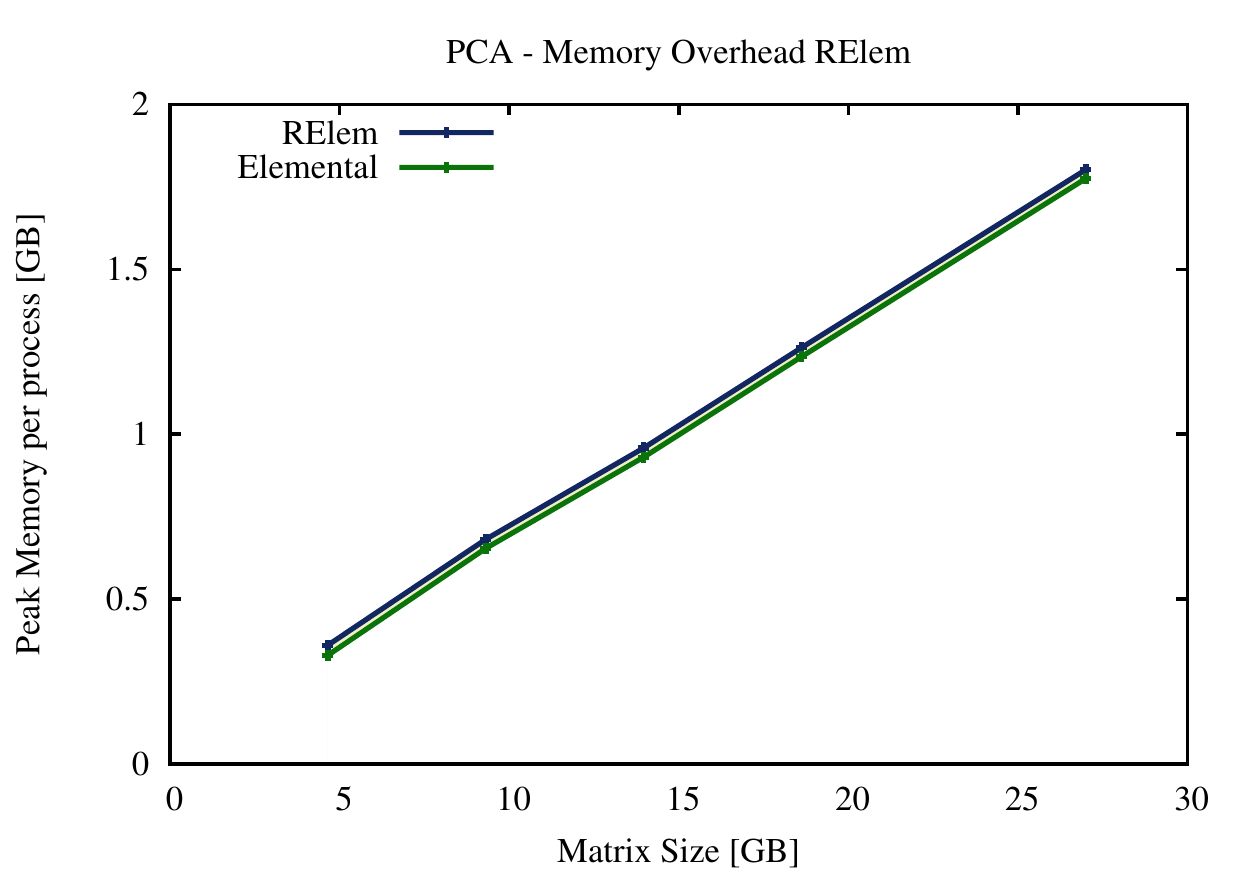}
    \caption{Memory requirements for distributed PCA with 72 processes.}
    \label{MemOverhead}
\end{figure}

Figure \ref{MemOverhead} presents the peak memory requirements of this
computation per process for both \relem and \elemental:  Independent of the
problem size, the memory overhead of \relem is constant at about 30MB per
process.  This corresponds to the memory footprint of the \R~environment and its
libraries, yet for large scale computations on gigabytes of data is negligible.

\section{Conclusions}
\label{conclusions}

The presented package \relem enables large scale distributed computations in~\R
through the state-of-the-art linear algebra library \elemental.  \relem allows
\R~users to tackle enormous problems on compute clusters that pure~\R cannot
possibly process being limited to a single shared memory processor.

By not only translating \elemental's interface to~\R but giving it a native~\R
feeling by overloading various operators and functions, we created a package
that both offers full access to \elemental for experienced users, and allows for
a seamless transition to distributed memory computations for \R~users.

Nonetheless, \relem is a light-weight interface from~\R to \elemental that in
both execution time and memory requirements introduces virtually no overhead at
all.  We demonstrated both \relem's performance and its scalability beyond the
memory-limitations of a single compute nodes.

\relem is available both on GitHub under the open-source MIT licence
(\url{https://github.com/HPAC/RElem}).

Future plans for \relem include the support for sparse linear algebra routines
available in \elemental.

\section*{Acknowledgments}
Financial support from The Aachen Institute for Advanced Study in Computational
Engineering Science (AICES) through the Deutsche Forschungsgemeinschaft (DFG)
grant GSC~111 is gratefully acknowledged. The authors thank Jack Poulson for his
technical support on Elemental.

\bibliography{article}

\end{document}